# Parallel transmit PUlse design for Saturation Homogeneity (PUSH) for Magnetization Transfer imaging at 7T

David Leitão[1,*], Raphael Tomi-Tricot[2], Pip Bridgen[1], Tom Wilkinson[1], Patrick Liebig[3], Rene Gumbrecht[3], Dieter Ritter[3], Sharon L. Giles[1], Ana Baburamani[4], Jan Sedlacik[1,4], Joseph V. Hajnal[1,4], and Shaihan J. Malik[1,4]

[1]Biomedical Engineering Department, School of Biomedical Engineering and Imaging Sciences, King's College London, London, United Kingdom

[2]MR Research Collaborations, Siemens Healthcare Limited, Frimley, United Kingdom

[3]Siemens Healthcare GmbH, Erlangen, Germany

[4]Centre for the Developing Brain, School of Biomedical Engineering and Imaging Sciences, King's College London, London, United Kingdom

[*]Corresponding author: Perinatal Imaging and Health, 1st Floor South Wing, St Thomas' Hospital, London SE1 7EH, UK. E-mail: david.leitao@kcl.ac.uk

## Abstract

**Purpose:** This work proposes a novel RF pulse design for parallel transmit (pTx) systems to obtain uniform saturation of semisolid magnetization for Magnetization Transfer (MT) contrast in the presence of transmit field ($B_1^+$) inhomogeneities. The semisolid magnetization is usually modeled as being purely longitudinal, with the applied $B_1^+$ field saturating but not rotating its magnetization, thus standard pTx pulse design methods do not apply.

**Theory and Methods:** Pulse design for Saturation Homogeneity (PUSH) optimizes pTx RF pulses by considering uniformity of root-mean squared $B_1^+$, $B_1^{rms}$, which relates to the rate of semisolid saturation. Here we considered designs consisting of a small number of spatially non-selective sub-pulses optimized over either a single 2D plane or 3D. Simulations and in vivo experiments on a 7T Terra system with an 8-TX Nova head coil in 5 subjects were carried out to study the homogenization of $B_1^{rms}$ and of the MT contrast by acquiring MT ratio maps.

**Results:** Simulations and in vivo experiments showed up to 6 and 2 times more uniform $B_1^{rms}$ compared to circular polarized (CP) mode for 2D and 3D optimizations, respectively. This translated into 4 and 1.25 times more uniform MT contrast, consistently for all subjects, where 2 sub-pulses were enough for the implementation and coil used.

**Conclusion:** The proposed PUSH method obtains more uniform and higher MT contrast than CP mode within the same SAR budget.

# 1. Introduction

Imaging at ultrahigh-field benefits from increased signal to noise ratio (SNR)[1] but is also hampered by larger transmit field ($B_1^+$) inhomogeneity. This can lead to non-uniform excitation of the magnetization and undesired spatially varying contrast. Various hardware[2,3] and pulse design[4,5] solutions have been proposed and one of the most flexible is parallel transmit[6–9] (pTx) which uses multiple transmit channels to enable spatial and temporal manipulation of the $B_1^+$ field. Its most basic form, 'static' $B_1^+$ shimming[10–12], attempts to create a more homogeneous $B_1^+$ field by applying amplitude and phase weightings to the individual transmit channels without changing the radiofrequency (RF) pulse waveforms themselves. Much greater control can be achieved if we instead consider the magnetization rotation using both RF and gradients; in this case a desired 'flip angle' distribution is usually designed by tailoring the RF pulse waveforms on each channel. This is often achieved using the small tip angle approximation in which case the design problem can be considered using the excitation k-space concept[13]. For larger rotations this concept breaks down but various methods still exist to design RF pulses by considering the rotation of magnetization directly[14–16].

Although a great diversity of RF pulse design methods exist, they typically have in common the use of the Bloch equation[17] to model the effect of the applied RF and/or gradient fields. Assuming short duration pulses, the effect of applying these fields is to rotate the magnetization. The Bloch equation can successfully model the magnetization dynamics of free water, but in biological tissues there is usually also a significant pool of semisolid magnetization[18,19]. This latter pool is affected differently by RF fields and can exchange magnetization with the free water, a phenomenon usually referred to as Magnetization Transfer[20,21] (MT). Two common assumptions of the semisolid pool are: (1) it can be modeled as having no transverse magnetization due to its very short $T_2^s \approx 10\mu s$; (2) its longitudinal magnetization directly saturates at a rate proportional to the applied RF power[22] (i.e. $|B_1^+|^2$). These properties and the coupling with free water magnetization can be modeled using the so-called binary-spin-bath (BSB) model[20].

Since the saturation of semisolid magnetization depends on $|B_1^+|^2$ and not simply $B_1^+$, the effect of $B_1^+$ inhomogeneities is more severe. Existing RF pulse design methods that might be used to correct for non-uniform flip angles in free water magnetization will fail if used for designing semisolid saturation pulses because: i) the semisolid has no transverse component that can be rotated by any applied gradients that are often used to improve excitation properties[23,24] and ii) the saturation rate of its longitudinal magnetization depends on $|B_1^+|^2$ and not $B_1^+$[20,22].

An important distinction here must be drawn between the saturation of semisolids, which is the subject of this work, and general 'saturation' pulses that are used (often with spoiler gradients) to suppress magnetization from free water and/or solutes. For the latter type there are examples using standard pTx pulse design methods[25,26], as there is transverse magnetization amenable to rotation from the RF pulses and gradients. For the remainder of this article the term 'saturation' is used to refer to semisolid saturation, unless otherwise specified.

In this work we propose a novel RF pulse design framework for semisolid saturation, called PUlse design for Saturation Homogeneity (PUSH). We first explore a general case of RF pulse design in the presence of semisolids, and then propose a simple exemplar method using trains of short sub-pulses with pTx and demonstrate the efficacy of this approach for MT-weighted imaging at 7T.

## 2. Theory
### 2.1 Physics models

The dynamics of free water ($f$) magnetization $\mathbf{M}^f = \begin{bmatrix} M_x^f & M_y^f & M_z^f \end{bmatrix}^T$ are described by the Bloch equation[17]:

$$\frac{d\mathbf{M}^f}{dt} = \left( \begin{bmatrix} 0 & \gamma \Delta B_z & -\gamma B_{1,y} \\ -\gamma \Delta B_z & 0 & \gamma B_{1,x} \\ \gamma B_{1,y} & -\gamma B_{1,x} & 0 \end{bmatrix} + \begin{bmatrix} -R_2^f & 0 & 0 \\ 0 & -R_2^f & 0 \\ 0 & 0 & -R_1^f \end{bmatrix} \right) \mathbf{M}^f + \begin{bmatrix} 0 \\ 0 \\ R_1^f M_0^f \end{bmatrix}$$

$$= (\mathbf{A} + \mathbf{E}) \mathbf{M}^f + \mathbf{c} \qquad [1]$$

where $\mathbf{A}$ comprises RF ($B_1^+ = B_{1,x} + iB_{1,y}$) and $B_0$ field variations ($\Delta B_z = \Delta B_0 + \mathbf{G} \cdot \mathbf{r}$) at coordinates $\mathbf{r}$ induced by off-resonance $\Delta B_0$ and gradients $\mathbf{G}$. Operators $\mathbf{E}$ and $\mathbf{c}$ contain relaxation effects through the relaxation rates $R_1^f$ ($= 1/T_1^f$) and $R_2^f$ ($= 1/T_2^f$), and $M_0^f$ is the equilibrium magnetization.

On the other hand, systems with MT can be described by the BSB model[22] which contains two pools corresponding to free water ($f$) and semisolid ($s$) magnetization:

$$\frac{d}{dt}\begin{bmatrix} \mathbf{M}^f \\ M_z^s \end{bmatrix} = \left( \begin{bmatrix} \mathbf{A} & 0 \\ 0 & -\langle W \rangle \end{bmatrix} + \begin{bmatrix} -R_2^f & 0 & 0 & 0 \\ 0 & -R_2^f & 0 & 0 \\ 0 & 0 & -k_{fs} - R_1^f & k_{sf} \\ 0 & 0 & k_{fs} & -k_{sf} - R_1^s \end{bmatrix} \right) \begin{bmatrix} \mathbf{M}^f \\ M_z^s \end{bmatrix} + \begin{bmatrix} \mathbf{c} \\ R_1^s M_0^s \end{bmatrix}$$

$$= (\tilde{\mathbf{A}} + \tilde{\mathbf{E}}) \begin{bmatrix} \mathbf{M}^f \\ M_z^s \end{bmatrix} + \tilde{\mathbf{c}} \qquad [2]$$

The operators $\tilde{\mathbf{A}}$, $\tilde{\mathbf{E}}$ and $\tilde{\mathbf{c}}$ include the effects of RF, gradients, relaxation, and exchange. In these expressions $R_1^s$ ($=1/T_1^s$) is the semisolid longitudinal relaxation rate, $M_0^s$ is the semisolid equilibrium magnetization, $k_{fs}$ is the exchange rate from $M_z^f$ to $M_z^s$ (vice-versa for $k_{sf}$), and $\langle W \rangle$ is the average saturation rate[22] that models the semisolid response to RF. In case of RF irradiation at a single off-resonance frequency $\omega$ then $\langle W \rangle$ is given by[22]

$$\langle W \rangle = \pi \gamma^2 g(\omega - \gamma \Delta B_z, T_2^s) \frac{1}{\tau} \int_0^\tau |B_1^+(t)|^2 dt = \pi \gamma^2 g(\omega - \gamma \Delta B_z, T_2^s) \langle |B_1^+|^2 \rangle \qquad [3]$$

where $\langle |B_1^+|^2 \rangle$ is the mean squared $B_1^+$ over pulse duration $\tau$, and $g$ is the semisolid absorption lineshape that depends on its transverse relaxation time $T_2^s$ and on the frequency shift $\omega - \gamma \Delta B_z$. Typically the absorption lineshape has much broader bandwidth[21] than the RF or $B_0$ field variations in the absence of gradients, such that $g(\omega - \gamma \Delta B_z, T_2^s) \approx g(\omega, T_2^s)$. It has also been observed that $g$ may have a chemical shift away from water (*e.g.*, Jiang et al[27] observed $\approx -2.6$ppm in white matter) – this shift should be considered part of the definition of $g$. Although Eq. [3] defines $\langle W \rangle$ for single frequency irradiation, it can also be calculated in some cases for RF pulses with multiple frequencies (e.g., multiband pulses[28]).

### 2.2 RF pulse design

In solving the Bloch equation (Eq. [1]) for a short RF pulse, matrix $\mathbf{E}$ and vector $\mathbf{c}$ can be neglected as relaxation typically occurs over a longer timescale, hence the magnetization dynamics comprises of rotations determined by $\mathbf{A}$. This can be solved by discretizing the sequence parameters in $N_t$ constant piecewise timesteps of duration $\Delta t$, each producing a rotation $\mathbf{R}(t)$:

$$\mathbf{M}^f(t + \Delta t) = \exp(\mathbf{A}(t) \Delta t) \mathbf{M}^f(t) = \mathbf{R}(t) \mathbf{M}^f(t) \qquad [4]$$

whereas the full rotation $\mathbf{R}^{full}$ of the magnetization can be calculated by taking left-wise multiplication over all $\mathbf{R}(t)$.

Similarly, in the BSB model (Eq. [2]) relaxation and exchange can be assumed to occur over a longer timescale than the typical RF pulse allowing them to be neglected for RF pulse design. The magnetization response is thus determined by $\widetilde{\mathbf{A}}$ and can also be solved by discretizing time:

$$\begin{bmatrix}\mathbf{M}^f\\ \mathbf{M}^s_z\end{bmatrix}(t+\Delta t) = \exp\left(\begin{bmatrix}\mathbf{A}(t) & 0\\ 0 & -\langle W\rangle\end{bmatrix}\Delta t\right)\begin{bmatrix}\mathbf{M}^f\\ \mathbf{M}^s_z\end{bmatrix}(t) \quad [5]$$

In the absence of exchange the response of the free water and semisolid pools is decoupled ($\widetilde{\mathbf{A}}$ is block diagonal), thus the matrix exponential is the exponential of its diagonal terms. The full magnetization response to an RF pulse can then be calculated by taking the product of the matrix exponentials from all $N_t$ timesteps over the duration $\tau$ of the RF:

$$\begin{bmatrix}\mathbf{M}^f\\ \mathbf{M}^s_z\end{bmatrix}(t+\tau) = \left(\prod_{t=1}^{N_t}\begin{bmatrix}\mathbf{R}(t) & 0\\ 0 & e^{-\langle W\rangle\Delta t}\end{bmatrix}\right)\begin{bmatrix}\mathbf{M}^f\\ \mathbf{M}^s_z\end{bmatrix}(t)$$

$$= \begin{bmatrix}\mathbf{R}^{\text{full}} & 0\\ 0 & e^{-\langle W\rangle\tau}\end{bmatrix}\begin{bmatrix}\mathbf{M}^f\\ \mathbf{M}^s_z\end{bmatrix}(t) \quad [6]$$

where the free water and semisolid pools responses are independent from each other. Design of RF pulses for free water magnetization usually target a desired flip angle $\alpha_{des}$ from rotation matrix $\mathbf{R}^{\text{full}}$. On the other hand, to control saturation of semisolid magnetization we can target a desired average saturation rate $\langle W\rangle_{des}$. Thus, the RF pulse design for both pools can be cast as a joint optimization:

$$\{\hat{\mathbf{b}},\hat{\mathbf{G}}\} := \arg\min_{\mathbf{b},\mathbf{G}}\{(1-\lambda)\|\alpha-\alpha_{des}\|_2^2 + \lambda\|\langle W\rangle - \langle W\rangle_{des}\|_2^2\} \quad [7]$$

where $\lambda \in [0,1]$ balances the error between the two terms and $\mathbf{b}$ and $\mathbf{G}$ are the RF and gradient waveforms, respectively.

## 2.3 Pulse design for Saturation Homogeneity

In this work we explored design of RF pulses to achieve uniform semisolid saturation, using pTx systems. To this end we have considered single frequency high power saturation pulses[29,30] (Figure 1A) applied at large offset frequency $\omega$. We assume that $\omega$ is much larger than the saturation pulse bandwidth such that we can neglect the effect on the free water magnetization, *i.e.*, it has null flip angle ($\alpha \approx 0$) – effectively performing the pulse design in Eq. [7] with $\lambda = 1$. Furthermore, according to Eq. [3] for single frequency irradiation we can control semisolid saturation using $\langle |B_1^+|^2\rangle$ instead of $\langle W\rangle$ in the pulse design.

For a pTx system the applied $B_1^+$ field is the linear superposition of the fields from its $N_{ch}$ channels:

$$B_1^+(\mathbf{r},t) = \sum_{j=1}^{N_{ch}} s_j(\mathbf{r})b_j(t) \quad [8]$$

where $s_j(\mathbf{r})$ are the transmit sensitivity maps (units of $\mu T/V$) and $b_j(t)$ are the RF waveforms for each channel (units of V). In this implementation we employed short repetition time (TR) sequences in which saturation depends on the cumulative effect over many TR periods, scaling with the mean squared $B_1^+$ (Eq. [3]) averaged over the TR[31,32] instead of the pulse duration $\tau$:

$$\langle |B_1^+|^2\rangle(\mathbf{r}) = \frac{1}{TR}\int_0^\tau |B_1^+(\mathbf{r},t)|^2 dt = \frac{1}{TR}\int_0^\tau \left|\sum_{j=1}^{N_{ch}} s_j(\mathbf{r})b_j(t)\right|^2 dt \quad [9]$$

Here the contribution from any other pulses during the same TR period (*e.g.*, excitation pulse as described in Methods) is neglected as they typically have much less power than the designed saturation pulse. The advantage of designing for $B_1^{rms}$ ($= \sqrt{\langle |B_1^+|^2\rangle}$) at the *sequence* rather than *pulse* level is that the former is typically the limiting factor for an MT-weighted sequence since it scales with the SAR; exposing this limit allows more flexibility to optimize the sequence within this constraint.

In this work we applied the same normalized waveform $b(t)$ (arbitrary units) in each channel scaled by a complex weight $w_j$ (units of V), moving the sum outside the integral:

$$\langle |B_1^+|^2 \rangle (\mathbf{r}, w_j) = \left| \sum_{j=1}^{N_{ch}} s_j(\mathbf{r}) w_j \right|^2 \frac{1}{TR} \int_0^\tau |b(t)|^2 dt = \left| \sum_{j=1}^{N_{ch}} s_j(\mathbf{r}) w_j \right|^2 \langle b^2 \rangle \quad [10]$$

where $\langle b^2 \rangle$ is the mean squared $B_1^+$ of the normalized waveform. Similarly to spokes[33,34]/$k_T$-points[35], the pulse can be extended by concatenating $N_{sp}$ sub-pulses, designated as PUSH-$N_{sp}$, with each sub-pulse weighted differently:

$$\langle |B_1^+|^2 \rangle (\mathbf{r}, w_{jp}) = \sum_{p=1}^{N_{sp}} \left| \sum_{j=1}^{N_{ch}} s_j(\mathbf{r}) w_{jp} \right|^2 \langle b^2 \rangle \quad [11]$$

where $p$ is the sub-pulse index. Each sub-pulse produces its own spatial mean squared $B_1^+$, such that the total $\langle |B_1^+|^2 \rangle$ is the sum of the contributions from all sub-pulses (example in Supporting Information Figure S1). Finally, the RF complex weights $w_{jp}$ can be designed to achieve a desired saturation by solving the optimization:

$$\widehat{w}_{jp} := \arg\min_{w_{jp}} \left\| \sqrt{\langle |B_1^+|^2 \rangle (\mathbf{r}, w_{jp})} - \beta(\mathbf{r}) \right\|_2^2$$

$$\text{s.t.} \quad \begin{aligned} & SAR_{10g,v} \leq SAR_{10g,max}, \quad 1 \leq v \leq N_{VOP}, \\ & \mathcal{P}_j(w_{jp}) \leq \mathcal{P}_{max}, \quad 1 \leq j \leq N_{ch}, \\ & |w_{jp}| \leq V_{max}, \quad 1 \leq p \leq N_{sp}, \quad 1 \leq j \leq N_{ch}. \end{aligned} \quad [12]$$

where $\beta(\mathbf{r})$ specifies the desired spatial $\sqrt{\langle |B_1^+|^2 \rangle} = \sqrt{\langle W \rangle / \pi \gamma^2 g(\omega, T_2^s)}$. Note that the optimization has been rewritten in terms of the square root of $\langle |B_1^+|^2 \rangle$ – doing so does not change the global optima. This was done so that the special case of a saturation pulse consisting of a single sub-pulse would reduce to Magnitude Least-Squares (MLS) $B_1^+$ shimming[11,12,36] of the saturation pulse. The terms 'PUSH-1 (1 sub-pulse)' and 'static shimming' will be used interchangeably from here onwards. Optimization was constrained to be within local SAR limits for a total of $N_{VOP}$ VOPs[37], as well as average power per channel $\mathcal{P}_j$ and maximum voltage $V_{max}$ per sub-pulse and per channel[38]. In the case of using the circular polarized (CP) mode, the complex weights are defined as $w_j = w_{CP} \exp(-i 2\pi(j-1)/N_{ch})$, where $w_{CP}$ was determined by minimizing the cost function (Eq. [12]) and $i^2 = -1$ denotes the imaginary unit.

## 3. Methods

All experiments were performed using a 7T scanner (MAGNETOM Terra, Siemens Healthcare, Erlangen, Germany) in prototype research configuration, with an 8Tx/32Rx head coil (Nova Medical, Wilmington MA, USA).

### 3.1 Pulse sequence setup

To illustrate the PUSH concept, we used a simple MT-weighted spoiled gradient-recalled echo (SPGR) sequence containing one saturation and one excitation pulse per TR (Figure 1A). The saturation sub-pulses (Figure 1B) applied at offset frequency $\omega = 2kHz$ had a Gaussian waveform (Time Bandwidth Product = 2.27, $\tau = 4ms$) and its complex weights were determined using Eq. [12]. Pulse optimization was solved in Matlab (Mathworks Inc., Natick, MA) using the interior-point algorithm from the *fmincon* routine, providing first and second order derivatives; constraints ($SAR_{10g,max} = 20W/kg$ over 8 VOPs provided by the vendor and in first level SAR mode[39], $V_{max} = 207V$, $\mathcal{P}_{max} = 24W$) were evaluated within the vendor pulse design framework included in the scanner console software (release Syngo.MR VE12U). A multi-start strategy with 10 random seeds proved to obtain consistent solutions.

As we focused on the saturation pulse design, the excitation was always in CP mode. To minimize the impact of the excitation pulse on the MT contrast, the flip angle was minimized balancing SNR and its inhomogeneity profile (Supporting Information Figure S2). This way the excitation pulse also had negligible power compared to the saturation pulse such that $\sqrt{\langle |B_1^+|^2 \rangle}$ in Eq. [12] can be assumed equivalent to the sequence $B_1^{rms}$.

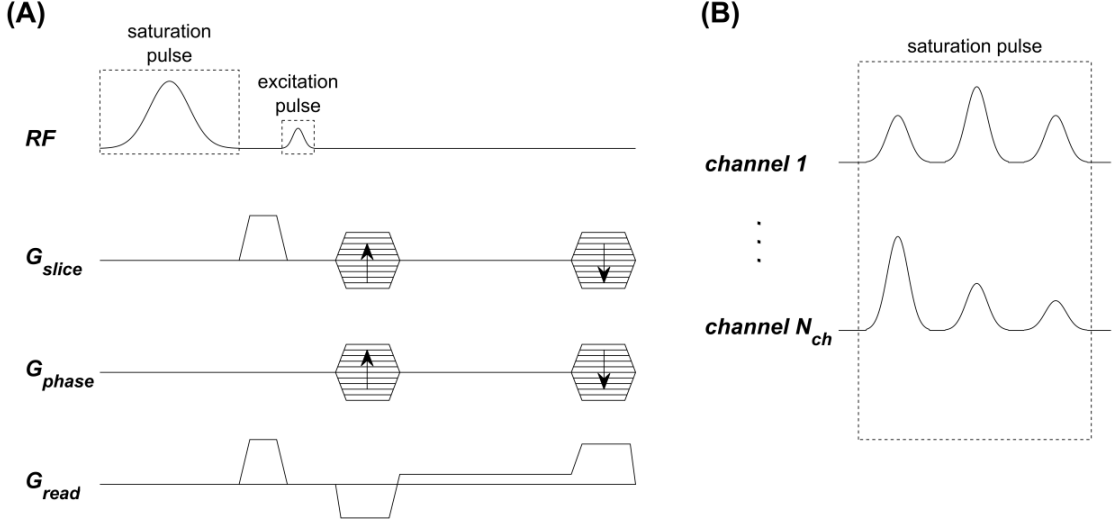

*Figure 1 - (A) Sequence diagram for one repetition time of the MT-weighted spoiled gradient-recalled-echo sequence used (here shown for 3D imaging but adaptable for 2D imaging). In this work the saturation pulse can be composed of several sub-pulses applied at large offset frequency as shown in (B) with 3 sub-pulses. These are individually scaled with complex weights per channel, where the weights are calculated with PUSH (Eq. [12]) or with CP mode. The excitation pulse is always applied in CP mode.*

## 3.2 Simulations

To explore the pulse design performance, saturation pulses with different number of sub-pulses (1, 2 and 3) were designed offline for a spatially invariant β ranging from $0.1\mu T$ to $2\mu T$ in steps of $0.1\mu T$. The optimizations were performed for both 2D axial slices and 3D volume of brain transmit maps from an 8-channel pTx system (details below), with each 2D single slice/3D optimization taking ≈ 7/22 seconds in Matlab R2020b (Mathworks Inc., Natick, MA) on a desktop computer (Intel i9-10900X @ 3.70GHz, 64GB of RAM, not parallelized). The solutions were analyzed in terms of their $B_1^{rms}$ maps and normalized root mean square error (NRMSE).

To predict the impact of spatial variation in mean squared $B_1^+$ on the MT contrast, Magnetization Transfer Ratio (MTR) maps were simulated using the definition:

$$\text{MTR}(\%) = 100 \times \frac{M_{ref} - M_{sat}}{M_{ref}} \quad [13]$$

where $M_{sat}$ and $M_{ref}$ are the steady-state signals acquired with and without the saturation pulse, respectively. For the MTR simulations the steady state of an SPGR sequence was calculated by solving Eq. [2] assuming the whole brain to have uniform tissue parameters similar to white matter[40,41]: $R_1^f = 0.4s^{-1}, T_2^f = 60\text{ms}, f = M_0^s/(M_0^s + M_0^f) = 0.1357, k = k_{sf}/(1-f) = k_{fs}/f = 32.79s^{-1}, R_1^s = 1.85s^{-1}, T_2^s = 9.6\mu s$ and a Super-Lorentzian absorption lineshape (centered at -773Hz)[27]. Different saturation pulses optimized offline for $\beta = 1\mu T$ were applied combined with a small excitation flip angle of $\alpha_{exc} = 5°$.

### 3.3 Experiments

In vivo scanning of five healthy volunteers was performed in accordance with local ethical approval. MTR maps were acquired for 2D and 3D imaging, as described in the subsections below. The saturation pulse was designed online as described in subsection 2.3 with calculation fully scanner-integrated within a Matlab R2012b (Mathworks Inc., Natick, MA) framework from the vendor, taking $\approx 30$ seconds. Prior to pulse design off-resonance $\Delta B_0$ mapping was performed using a dual-echo SPGR sequence and $B_1^+$ mapping was performed using a pre-saturation turbo-FLASH sequence[42] (both with resolution $4 \times 4 \times 6$ mm$^3$); $B_1^+$ was corrected for bias using an empirically determined correction factor[43]. The reference voltage ($V_{ref}$) determined by the scanner's built-in adjustment steps was also recorded: a higher reference $V_{ref}$ indicates lower efficiency in generating $B_1^+$, and hence lower achieved $B_1^+$ for a given SAR level. A signal intensity-based mask from the vendor's framework was used to not impair the workflow but was pre-processed to remove non-brain tissue voxels by eroding each axial slice with a 3-pixel (12mm) radius disk and cropping axial slices that included voxels from the mouth and jaw.

Prior to MTR maps calculation (Eq. [13]), images were registered together using FSL BET[44] and FSL FLIRT[45], and white matter (WM) segmentation was performed using FSL FAST[46], further eroded with a 1-pixel radius disk to reduce partial volume effects. $B_1^{rms}$ maps were simulated retrospectively using the $B_1^+$ maps and the pulses optimized online (during the scan).

#### 3.3.1 2D imaging

2D MTR maps were acquired in all subjects for a single axial slice in the middle of the brain (resolution $1 \times 1 \times 5$ mm$^3$, matrix size $220 \times 220$, TR=22ms, TE=4ms, BW = 220Hz/Px, 4 averages). Data were acquired using three different saturation pulses (CP mode, PUSH-1, PUSH-2) and for four $\beta$ ($0.7\mu T, 1.0\mu T, 1.3\mu T, 1.6\mu T$). For each MTR map a set of $M_{sat}$ and $M_{ref}$ images were acquired, firstly $M_{sat}$ with 30 seconds of dummy pulses (to stabilize the RF output) followed immediately by $M_{ref}$ with 10 seconds of dummy pulses, resulting in $T_{acq} = 1:19$s per MTR map. Fewer dummy cycles were required for $M_{ref}$ since it was acquired immediately after $M_{sat}$. White matter segmentation for further analysis was performed using the MTR map obtained with PUSH-2 at $\beta = 1.3\mu T$ due to its uniform contrast (as shown later).

#### 3.3.2 3D imaging

3D whole brain MTR maps (resolution $1 \times 1 \times 1$ mm$^3$, matrix size $220 \times 220 \times 176$, TR=22ms, TE=4ms, BW = 220Hz/Px, GRAPPA[47] acceleration factor of $2 \times 2$ and elliptical shutter) were acquired in two subjects with $\beta = 1\mu T$ using three different saturation pulses (CP mode, PUSH-1, PUSH-2). A single $M_{ref}$ volume was acquired plus three $M_{sat}$ volumes, one for each saturation pulse. All volumes were acquired with 30 seconds of dummy pulses, resulting in $T_{acq} = 4:26$s per volume. An additional MP2RAGE[48] acquired at the same resolution was used for segmentation.

#### 3.3.3 Gradient blip experiment

To explore the impact of applying gradients between RF sub-pulses on the semisolid saturation, we carried an experiment adding gradient blips to the PUSH saturation pulse trains. These gradients would affect the 'flip angle' if this pulse were applied to free water magnetization, but not the semisolid saturation if it behaves as modeled with longitudinal magnetization only. 2D MTR maps (same protocol as in 3.3.1 with TR=27ms) were acquired in one subject with saturation pulses designed using PUSH-3 for $\beta = 1\mu T$, both excluding and including gradient blips between the sub-pulses. For the latter the gradient blips were $100\mu s$ in duration in the x and y-directions, each producing a $4\pi$ phase roll across the FOV. The $B_1^{rms}$ and flip angle maps of each pulse were computed (flip angle calculated at the RF offset frequency with Eq. [6]) and compared to the measured MTR maps.

## 4. Results

### 4.1 Simulations

Figure 2 examines the error in $B_1^{rms}$ as a function of the target β for 2D optimization on a middle axial slice (Figure 2A) and 3D whole-brain optimization (Figure 2B). Supporting Information Figure S3 further expands on Figure 2A for other axial slices. In all cases CP mode has a constant NRMSE until reaching the local SAR limits when the error increases as the voltage is capped. On the other hand, PUSH-1, -2 and -3 perform better than CP mode across all β for both 2D and 3D imaging. For the 2D case PUSH-1 (i.e. static shimming) gives an NRMSE 2 times smaller than CP mode for $\beta \leq 0.4\mu T$ in the middle slice (Figure 2A) but its performance worsens as β increases. PUSH-2 and -3 perform equally well, with a constant NRMSE 6 times smaller than CP mode (for $\beta \leq 1\mu T$) in the middle slice. Remarkably, PUSH-2 and -3 still perform well for $\beta \geq 1\mu T$, beyond where CP mode reached the local SAR limits. Generally, in 2D the performance gain offered by PUSH was larger for middle and inferior slices while superior slices had less inhomogeneity to begin with.

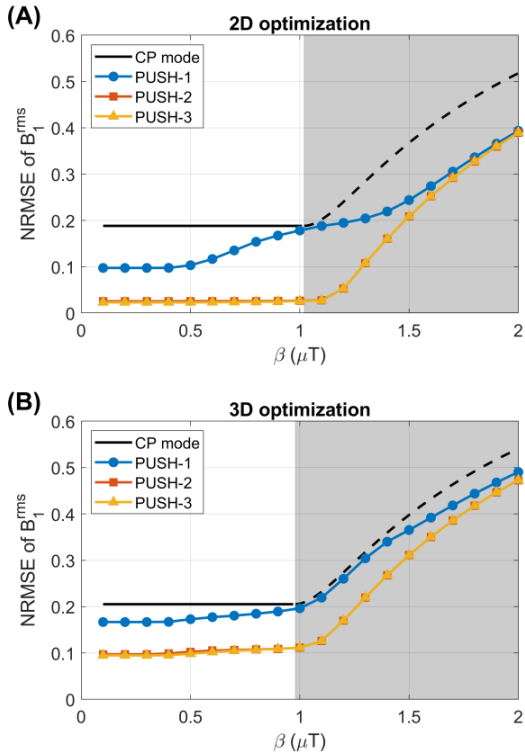

*Figure 2 - NRMSE of the $B_1^{rms}$ for (A) 2D middle axial slice and (B) 3D volume, comparing CP mode with optimized PUSH solutions using 1, 2 and 3 sub-pulses (curves for 2 and 3 sub-pulses are superimposed due to nearly identical performance). The gray area represents β where CP mode reached the local SAR limits and its voltage capped.*

For 3D imaging (Figure 2B) similar trends were observed, but with overall larger NRMSE. For β ≤ 1µT PUSH-1 obtains an NRMSE ≈ 17% smaller than CP mode, whereas PUSH-2 and -3 perform again equally well and still achieve an NRMSE 2 times smaller than CP mode.

Figures 3 and 4 show computed $B_1^{rms}$ maps for 2D and 3D imaging, respectively, for some of the solutions in Figure 2. Note that in Figure 4 some stripes are visible in the sagittal and coronal planes; these were found to be caused by artefacts in the acquired $B_1^+$ maps. In Figure 3 the $B_1^{rms}$ for CP mode scales up with β and caps after reaching the local SAR limits. PUSH-1 achieves more uniform $B_1^{rms}$ for β ≤ 0.7µT but gets progressively less homogeneous with increasing β, producing solutions with "holes". PUSH-2 and -3 achieve $B_1^{rms}$ maps similar to one another that are more uniform up to larger β. For 3D imaging, Figure 4 shows that CP mode produces a $B_1^{rms}$ pattern with center brightening. The $B_1^{rms}$ produced by PUSH-1 is very similar to CP mode with slight improvements for β ≤ 0.7µT. With PUSH-2 and -3 the $B_1^{rms}$ is more uniform in the middle slices and up to larger β, however both solutions underdeliver in the superior and inferior slices of the brain.

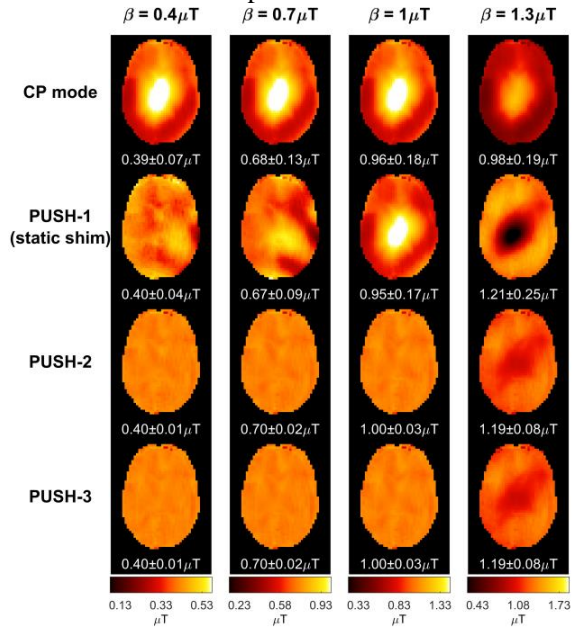

*Figure 3 – 2D $B_1^{rms}$ maps for some solutions in Figure 2. Columns contain maps for different β, increasing from left to right. Rows contain different saturation pulses: CP mode (top), PUSH-1 - i.e. static shimming - (second from top), PUSH-2 (second from bottom) and PUSH-3 (bottom). For each combination the mean ± standard deviation of $B_1^{rms}$ is shown below the respective maps.*

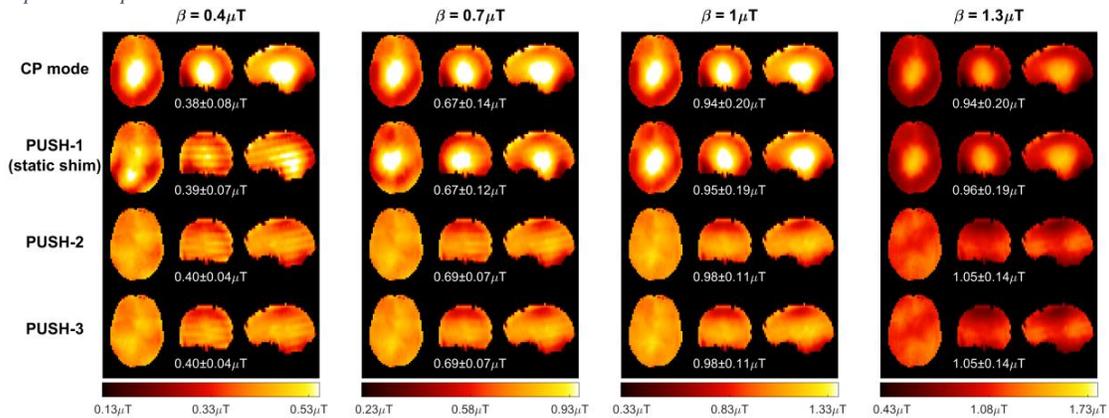

*Figure 4 – Transverse, coronal and sagittal planes of the 3D $B_1^{rms}$ maps for some β and solutions in Figure 2. Columns contain maps for different β, increasing from left to right. Rows contain different saturation pulses: CP mode (top), PUSH-1*

*- i.e. static shimming - (second from top), PUSH-2 (second from bottom) and PUSH-3 (bottom). For each combination the mean ± standard deviation of $B_1^{rms}$ is shown below the respective maps.*

Figure 5 shows simulated 2D and 3D MTR maps for β=1μT solutions from Figure 2. Note that all simulated MTR maps have been calculated using white matter properties over the whole brain: they are intended to visualize the spatial variations of saturation rather than the actual MTR contrast that will be seen in a scan. Both CP mode and PUSH-1 solutions yield similar MTR maps with center brightening for both 2D and 3D. However, for 2D imaging PUSH-1 can also yield solutions with contrast "holes" for some slices. On the other hand, PUSH-2 and -3 achieve similar strong improvement in 2D and 3D, with a standard deviation ≈ 2 times smaller than CP mode. A drop in achieved MTR towards the superior and inferior slices is seen in 3D. For all cases the MTR maps have a good correlation with the respective $B_1^{rms}$ in Figures 3 and 4.

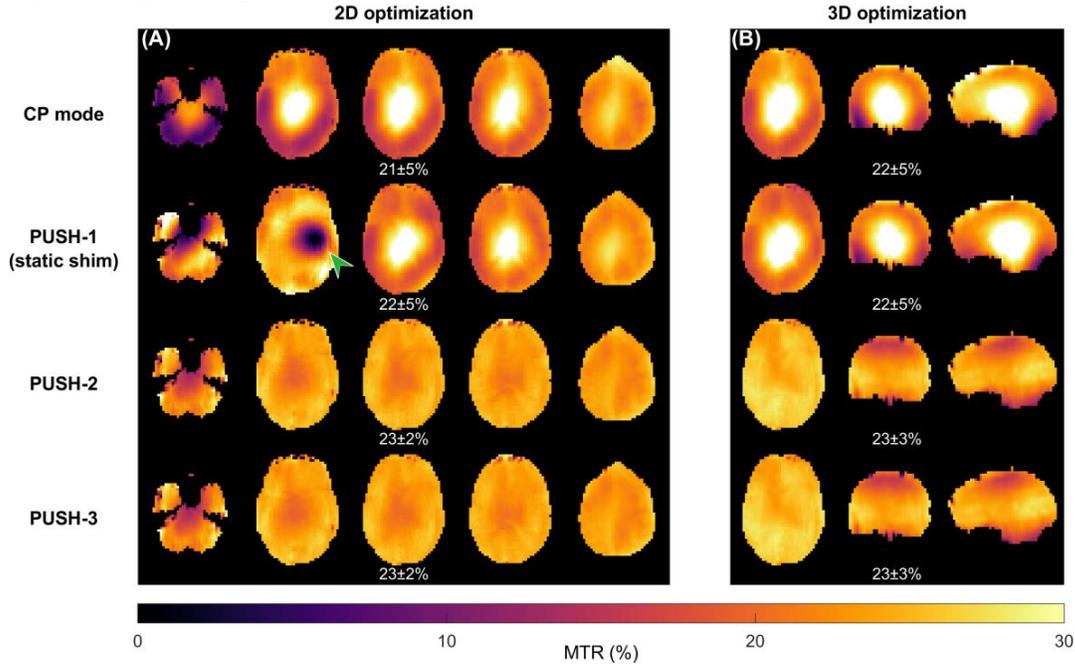

*Figure 5 - Simulated (A) 2D and (B) 3D MTR maps using solutions (CP mode, PUSH-1, PUSH-2 and PUSH-3) from Figure 2 for β=1μT as the saturation pulses. For 2D slices 6, 10, 12 (slice in Figure 3), 14 and 18 (from left to right) that were individually optimized are shown. The green arrow in (A) points to a contrast "hole" seen in some slices optimized with PUSH-1. The mean ± standard deviation of MTR over the whole volume is shown below the respective maps.*

### 4.2 Experiments
### 4.2.1 2D imaging

2D MTR maps from one subject are shown in Figure 6 for different pulses and β. CP mode shows a constant contrast pattern brighter in the center, that scales up with the β until it reaches the local SAR limits, after which the voltage is capped and no more RF power is delivered with increasing β. PUSH-1 achieves more uniform contrast for the smallest β but then resembles CP mode up to the largest β where it has a "hole" in the contrast whilst increasing the contrast everywhere else. PUSH-2 yields uniform contrast for all β with up to 4 times smaller dispersion except for $β = 1.6μT$, where it is slightly less bright in the center. The MTR maps correlate very well with the corresponding $B_1^{rms}$ maps in Supporting Information Figure S4.

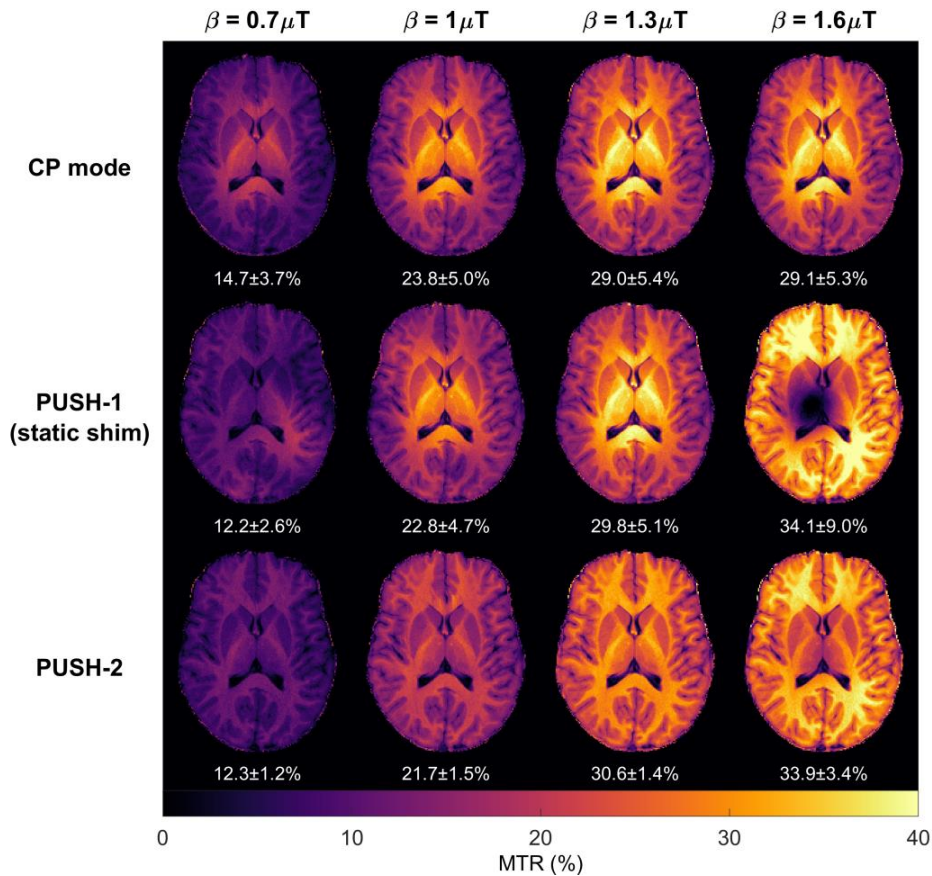

*Figure 6 - 2D MTR maps from one subject (subject A in Figure 7) using different saturation pulses (top row: CP mode; middle row: PUSH-1; bottom row: PUSH-2) and different β (across the columns, increasing from left to right). Below each is the mean ± standard deviation of MTR over the white matter mask.*

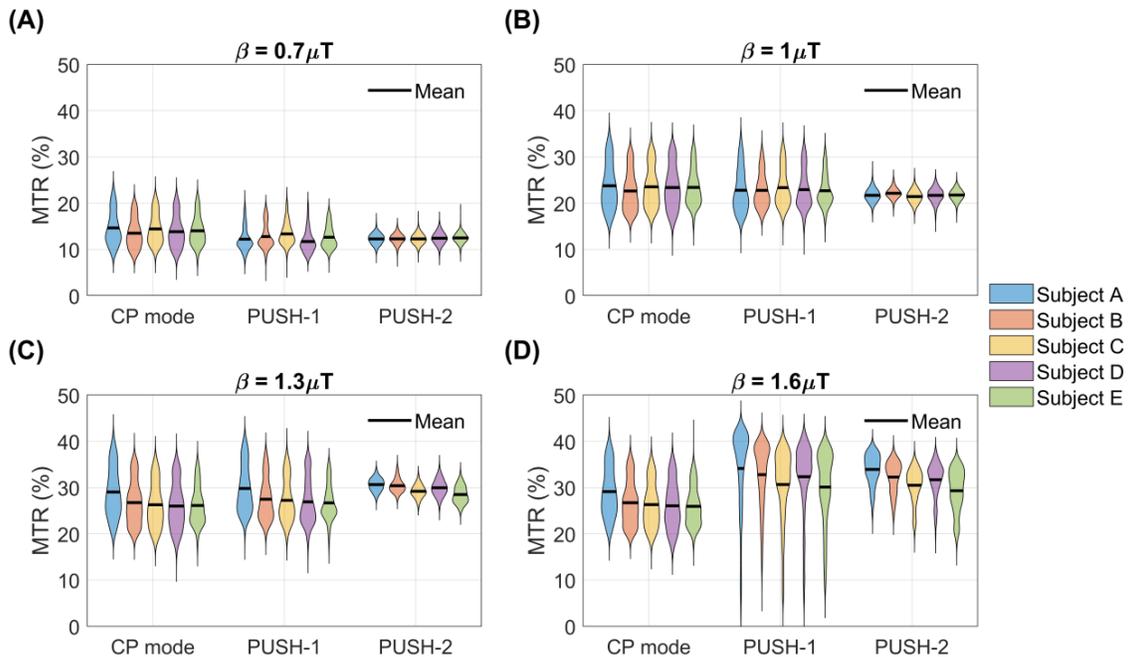

*Figure 7 - Violin plots of the MTR distribution in white matter for all 5 subjects scanned with the 2D imaging protocol. MTR distributions for saturation pulses designed using β of (A) 0.7μT, (B) 1μT, (C) 1.3μT and (D) 1.6μT. The black line represents*

*the mean MTR. Subjects are sorted in increasing order of reference voltage ($V_{ref} = \{245, 254, 259, 260, 272\}V$), which is inversely proportional to the maximum β achievable with CP mode.*

Figure 7 illustrates the MTR distribution in white matter for 5 subjects as a function of the saturation pulse and β; the subjects are arranged in order of increasing $V_{ref}$. The MTR distributions are consistent across all subjects, with PUSH-2 yielding narrower distributions for all β. PUSH-1 has narrower distributions for the smallest β, but at $1.6\mu T$ its distribution exhibits a heavy tail towards low MTR values as its maps have "holes" (Figure 6). Nevertheless, both PUSH-1 and -2 can achieve higher mean MTR than CP mode for the largest β, as in all subjects CP mode reached the local SAR limits below $1.3\mu T$. The MTR in Figure 7 correlates well with the corresponding $B_1^{rms}$ distributions in Supporting Information Figure S5. The subjects are ordered by increasing $V_{ref}$, showing an expected negative correlation between MTR and $V_{ref}$ for the largest β (also illustrated by Supporting Information Figure S6), as higher $V_{ref}$ means lower maximum $B_1^+$ peak.

### 4.2.2 3D imaging

Figure 8 shows the 3D MTR maps ($\beta = 1\mu T$) and their distribution in WM for one subject. CP mode and PUSH-1 show similar contrast with center brightening. On the other hand, PUSH-2 yields approximately 25% more uniform MTR (standard deviation over WM mask), as indicated by the narrower and taller histograms, especially in the middle and bottom slabs. The MTR maps and histograms correlate well with the corresponding $B_1^{rms}$ in Supporting Information Figure S7. MTR and $B_1^{rms}$ maps for a second subject are given in Supporting Information Figures S8 and S9, which show similar results.

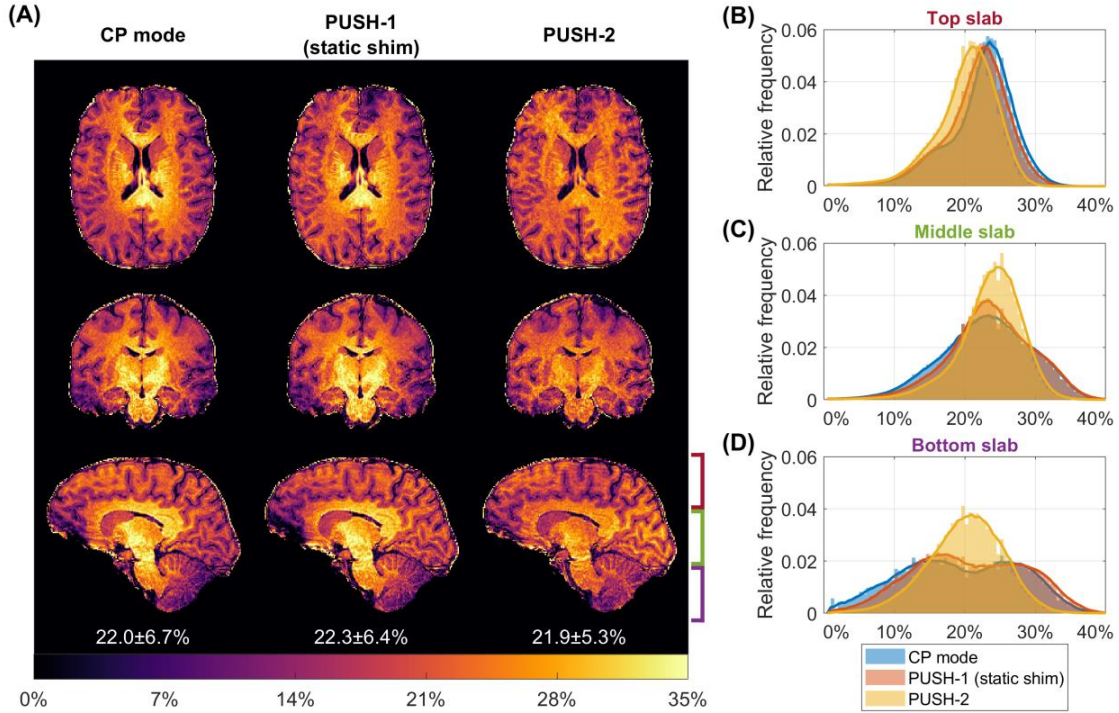

*Figure 8 – (A) Transverse, coronal and sagittal planes of the 3D MTR maps for subject C ($\beta = 1\mu T$). The left column contains the MTR maps acquired using CP mode, middle column using PUSH-1 and right column using PUSH-2. Below each sagittal plane is the mean ± standard deviation of MTR over the white matter mask. (B-D) Histograms of the MTR distribution in white matter over three slabs: (B) top slab, (C) middle slab, and (D) bottom slab, as illustrated in (A) near the bottom right sagittal plane. Moving average plotted jointly with histograms to delineate distribution trend.*

### 4.2.3 Gradient blip experiment

Figure 9 shows results from PUSH-3 pulses without (Figure 9A) and with (Figure 9E) gradient blips between sub-pulses; MTR maps were acquired while $B_1^{rms}$ and flip angle maps were simulated from acquired $B_1^+$ and $\Delta B_0$ maps. The MTR maps (Figure 9D and 9H) are virtually identical and very uniform, in agreement with their respective $B_1^{rms}$ maps (Figures 9B and 9F). The flip angle maps (Figures 9C and 9G) computed by also considering rotation induced by the gradients have a very different appearance and are both non-uniform.

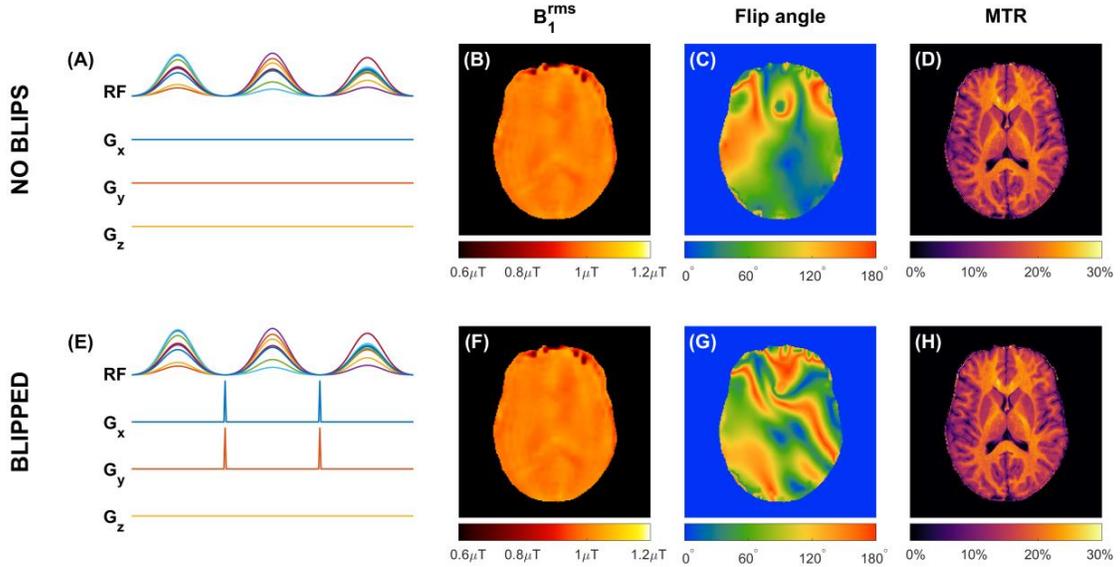

*Figure 9 – Experiment using the same PUSH-3 pulses without (A-D) and with (E-H) gradient blips between sub-pulses, each blip producing a dephasing of $4\pi$ across the x- and y-FOV. Corresponding (B,F) $B_1^{rms}$, (C,G) flip angle at the RF offset frequency and (D,H) MTR maps.*

## 5. Discussion

This work presents a novel pTx pulse design to overcome $B_1^+$ inhomogeneity in MT imaging at ultrahigh field by controlling semisolid saturation through the mean squared $B_1^+$. This can be performed either instead of or in addition to controlling the excitation properties of water for a given RF pulse; in this work we focused on testing the former case which is relevant to application of off-resonance saturation pulses. PUSH was tested in simulations and in vivo, yielding more uniform MT contrast.

Current pTx pulse design methods[24,33–35] usually employ a combination of RF and $B_0$ gradients to optimize the flip angle of the resulting excitation. These methods are unsuitable for designing semisolid saturation pulses for MT imaging since the semisolid pool has no transverse magnetization and instead saturates directly with $|B_1^+|^2$. This was experimentally confirmed (Figure 9) by applying the same RF pulse twice, with and without gradient blips in-between its sub-pulses. The gradients blips do not change $|B_1^+|^2$ but drastically alter the flip angle if applied to free water magnetization. The measured MTR maps show no difference in the MT contrast, supporting the fact that 'flip angle' is not a useful metric to use when designing or describing semisolid saturation pulses.

A simpler alternative to 'dynamic pTx' pulse design is $B_1^+$ shimming[11,12,36] which aims to create a spatially uniform $B_1^+$ distribution. An optimal $B_1^+$ shimming solution would also achieve a uniform $|B_1^+|^2$ meaning that in principle $B_1^+$ shimming is a special case of PUSH where the RF pulse is 'static' (i.e. pTx degrees of freedom are not modulated through the pulse). To connect PUSH with $B_1^+$ shimming, the optimization (Eq. [12]) is formulated in terms of the square-root of $\langle |B_1^+|^2 \rangle$, such that with 1 sub-pulse it simplifies to a magnitude least-squares[11,12,36] $B_1^+$ shimming design.

A related alternative approach is the MIMOSA[49] pTx design proposed to homogenize saturation in pulsed CEST. In that case a train of saturation pulses interleaved with spoiling gradients can be approximated by an equivalent continuous wave saturation whose effective $B_1^+$ is better described by the $B_1^{rms}$[50] over the pulse train. Hence the MIMOSA design applies two complementary modes[51,52] with the objective of homogenizing $B_1^{rms}$ for CEST saturation. This effectively results in a solution equivalent to PUSH-2, but for the special case of using two pre-defined modes of the pTx coil.

## 5.1 PUSH performance for MTR imaging at 7T

Simulations show that in both 2D and 3D imaging (Figure 2) PUSH-2 and -3 yield a strong improvement in the homogeneity of $B_1^{rms}$ for all β achievable (within SAR limits) with CP mode, whereas with PUSH-1 (static shimming) the improvements reduce as β increases. Remarkably, PUSH-2 and -3 sustain these improvements beyond β achievable with CP mode, delivering higher and more uniform $B_1^{rms}$. For 2D imaging (Figure 3 and Supporting Information Figure S3) these improvements are more substantial in the inferior and middle axial slices of the brain, however the maximum β achievable with CP mode is considerably smaller for the inferior slices. For 3D imaging (Figure 4) these improvements are smaller, with $B_1^{rms}$ decreasing from the middle towards the inferior and superior slices. In all simulations 2 or 3 sub-pulses yield very similar results, suggesting that 2 sub-pulses are enough to explore the variability in the transmit sensitivity maps used, hence in most in vivo experiments a maximum of 2 sub-pulses was used.

The 2D in vivo experiments shows up to 4 times more uniform MTR maps with PUSH-2 (Figure 6), corroborated by the narrow distributions of MTR in WM (Figure 7). Higher maximum MTR is obtained with PUSH-2, as expected from the simulations. Moreover, PUSH-1 (static shimming) solutions for $β = 1.6 \mu T$ have pathological contrast with "holes", which is known to affect shimming solutions[53]. The 2D results are consistent across all subjects, with the MTR maps correlating very well with the respective $B_1^{rms}$ maps. Some inter-subject variability is observed for the maximum MTR achieved (Figure 7D), relating to the reference voltage of each subject (Supporting Information Figure S6). This is understandable since a higher reference voltage indicates the subject experiences a higher SAR per unit of achieved $B_1^+$ leaving less room for optimization.

The 3D in vivo experiments also show more homogeneous MTR maps with PUSH-2 (Figure 8) but with a more modest 25% improvement in homogeneity. The distribution in WM shows smaller MTR values in the top and bottom slabs, agreeing with the respective $B_1^{rms}$ distribution (Supporting Information Figure S7). This effect is also observed in the 3D simulations (Figure 4). A potential half-way point between the 2D and 3D results would be to use a multi-slab approach where saturation pulses are designed separately for each slab (though they are spatially non-selective due to the broad semisolid lineshape) and are paired with slab selective excitation pulses.

## 5.2 Impact of RF coil design

While current pTx methods can employ gradients to enhance spatial encoding of RF pulses designed to achieve rotations of magnetization, PUSH relies solely on the transmit sensitivity maps to homogenize the MT contrast. This is seen particularly in the performance of PUSH in 3D where there is a persistent decrease in the achieved MTR in the superior and inferior regions. This is consistent with the limited coverage and lack of pTx control over $B_1^+$ variation in the z-axis (head-foot) from the circumferentially arranged transmit elements in the coil used. It is likely that the proposed method would benefit from alternative coil geometries[10], *e.g.* more channels and/or different distribution, to achieve a greater control of the mean squared $B_1^+$ spatial distribution.

For the coil used in this work we found that more than 2 sub-pulses does not improve the MT contrast homogeneity, but this may also prove not to be the case for alternative coil designs. More sub-pulses might also be beneficial in the case where peak voltage is the active constraint, whereas in the current implementation with the sequence and hardware used, local SAR was always the most limiting.

### 5.3 Assumptions and future extensions

Although exchange (Eq. [2]) is neglected over the RF duration, its cumulative effect over the whole pulse sequence makes MTR sensitive to both saturation of the semisolid and rotation of the free water[54,55]. Thus the excitation pulse can also affect MT contrast because i) it applies some power and ii) it rotates the free water magnetization that exchanges with the semisolid pool. In order to focus only on the saturation pulses our experiments employed a low excitation flip angle to minimize MTR sensitivity to excitation inhomogeneities, as excitation pulses used CP mode. As a result, the observed MTR is highly correlated with the $B_1^{rms}$ over the entire sequence; the contribution of excitation pulses to the $B_1^{rms}$ is negligible. This simple embodiment is used as a means to illustrate the key concept; however a future implementation might also consider designing uniform excitation pulses using methods such as k$_T$-points[35] or SPINS[24] potentially as part of a joint optimization problem (Eq. [7]). Likewise, it is not necessary to compute pulses in terms of the sequence $B_1^{rms}$ as done here. Use of the root-mean squared instead mean squared $B_1^+$ in Eq. [12] has the advantage that in the case of 1 sub-pulse it simplifies to a static shimming problem, making it a special case and allowing for a direct comparison. Likewise, averaging over the sequence TR rather than the pulse duration connects more closely to the expected SAR limits[31], and MT contrast for sequences with short TR where the continuous wave approximation is still valid[31,32]. However, in sequences with long TR this approximation breaks down and different exchange times affect MT contrast, so it is more appropriate to consider $\langle |B_1^+|^2 \rangle$ over the pulse duration (as in Eq. [3]) instead of over the TR (Eq. [9]).

Although gradient blips are observed not to affect the MT contrast, gradients applied *during* (as opposed to in between) RF pulses are expected to affect the semisolid saturation. According to Eq. [3] the saturation depends on $\langle |B_1^+|^2 \rangle$ but also on the absorption lineshape $g(\omega - \gamma \Delta B_z, T_2^s)$. Thus, theoretically it is also possible to control the saturation using applied gradients, which could be an avenue to explore, though this would require prior knowledge of the absorption lineshape[21,56].

## 6. Conclusion

This work proposed a novel RF pulse framework called PUlse design for Saturation Homogeneity (PUSH) for design of RF pulses considering their saturation effect on semisolid magnetization relevant to magnetization transfer imaging. It was also demonstrated that adding gradient blips between RF sub pulses as commonly used by standard pTx methods does not affect the MT contrast; the 'flip angle' of a saturation pulse is not a meaningful way of describing its operation.

The specific case demonstrated in this work was the design of off-resonance saturation pulses where on-resonance effects can be neglected. Simulations and in vivo experiments showed that for the 8 channel RF coil used in this work PUSH can obtain up to 4 and 1.25 times more uniform MT contrast in 2D and 3D imaging, respectively, achieving monomodal distributions of MTR that correlate very well with the corresponding applied $B_1^{rms}$. Moreover, PUSH delivered higher $B_1^{rms}$ than CP mode under the same SAR budget, thus also obtaining stronger contrast.

# Acknowledgments

This research was funded in whole, or in part, by the Wellcome Trust [WT 203148/Z/16/Z]. For the purpose of open access, the author has applied a CC BY public copyright licence to any Author Accepted Manuscript version arising from this submission. This work was supported by the Wellcome/EPSRC Centre for Medical Engineering [WT 203148/Z/16/Z] and by the National Institute for Health Research (NIHR) Biomedical Research Centre based at Guy's and St Thomas' NHS Foundation Trust and King's College London and/or the NIHR Clinical Research Facility, and funded by the King's College London & Imperial College London EPSRC Centre for Doctoral Training in Medical Imaging [EP/L015226/1]. The views expressed are those of the author(s) and not necessarily those of the NHS, the NIHR or the Department of Health and Social Care.

# Data availability statement

According to UK research councils' Common Principles on Data Policy and Wellcome Trust's Policy on data, software and materials management and sharing, all simulated data supporting this study will be openly available at https://github.com/mriphysics/PUSH (hash 3ce6d8a was the version at time of submission). This excludes in-vivo MRI data because of the terms of the ethical approval under which they were acquired, and proprietary code from SIEMENS but that can be shared upon request by agreement including the vendor.

# Parallel transmit PUlse design for Saturation Homogeneity (PUSH) for Magnetization Transfer imaging at 7T: Supporting Information

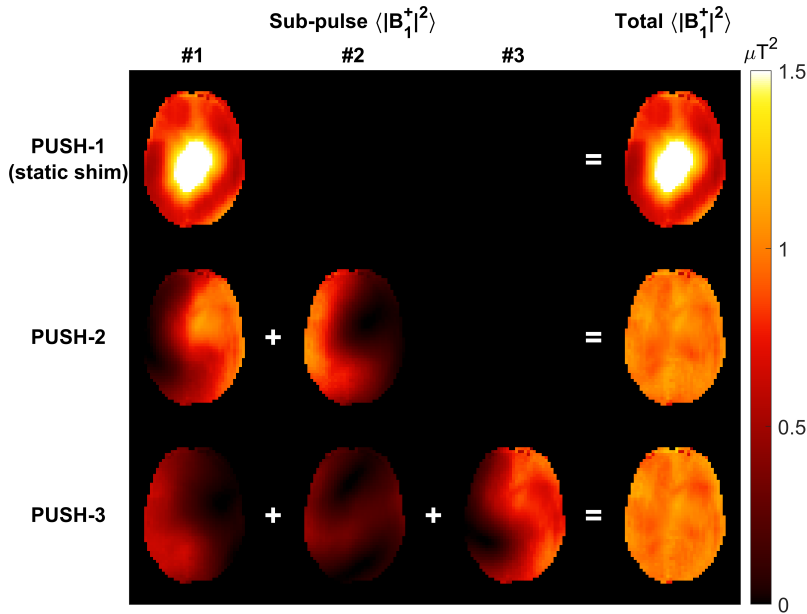

Supporting Information Figure S1: Mean squared $B_1^+$ ($\langle |B_1^+|^2 \rangle = \beta^2$) of each sub-pulse for the PUSH-1, -2 and -3 pulses optimized in the Simulations section 3.2 (middle axial slice, $\beta = 1\mu T$). The first three columns show the sub-pulse $\langle |B_1^+|^2 \rangle$, whereas the last column shows the total $\langle |B_1^+|^2 \rangle$ which is the sum of the contributions from all sub-pulses. For PUSH-2 and -3 the sub-pulses are highly complementary, yielding very uniform total $\langle |B_1^+|^2 \rangle$.

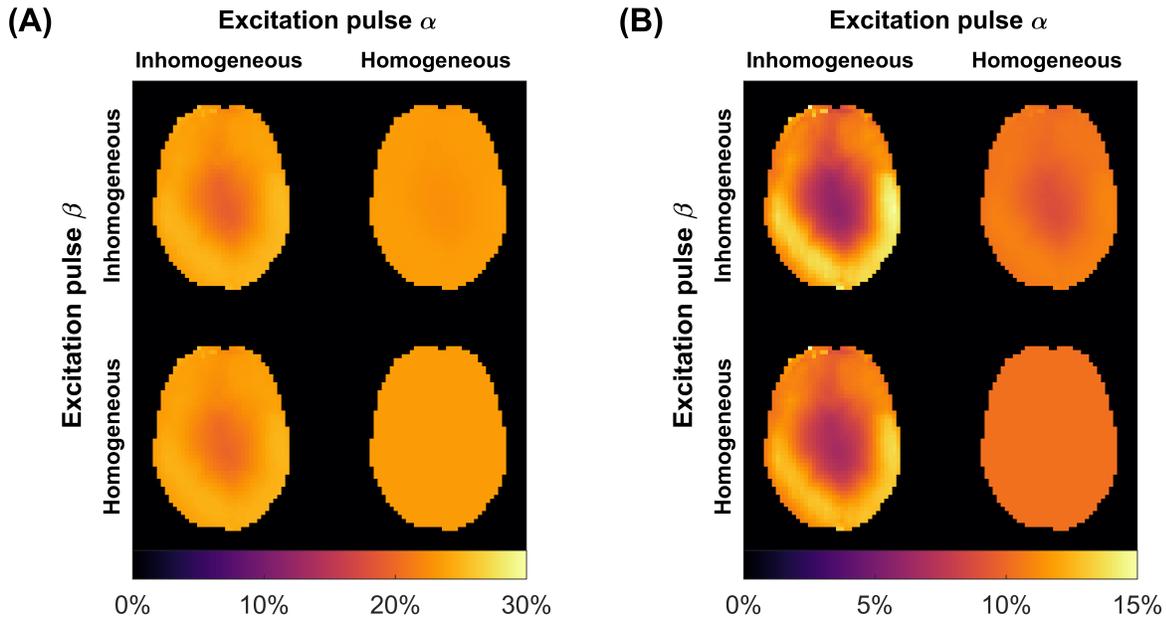

Supporting Information Figure S2: MTR simulations assuming an ideal homogeneous saturation pulse whilst alternately changing $\alpha$ and $\beta$ properties of the excitation pulse from spatially inhomogeneous to homogeneous. Simulations using target flip angle of (A) 5° and (B) 15°, showing that for 5° any inhomogeneity in either $\alpha$ or $\beta$ induces small changes in MTR, whereas for 15° the induced changes are much larger, with $\alpha$ inhomogeneity being the largest confound. In these simulations the pattern from CP mode was used for the inhomogeneous profiles.

Supporting Information Figure S3: (A) NRMSE of $B_1^{rms}$ for the axial slice positioned as indicated by the red line in the sagittal plane in (B), comparing CP mode with the optimized PUSH solutions using 1, 2 and 3 sub-pulses (curves for 2 and 3 sub-pulses are superimposed due to nearly identical performance). The gray area represents $\beta$ where CP mode reached the local SAR limits and its voltage is capped. Slice 12 corresponds to the solution in Figure 2A. To navigate through different slices this document needs to be open on a JavaScript-supporting PDF viewer, such as Adobe Acrobat Reader.

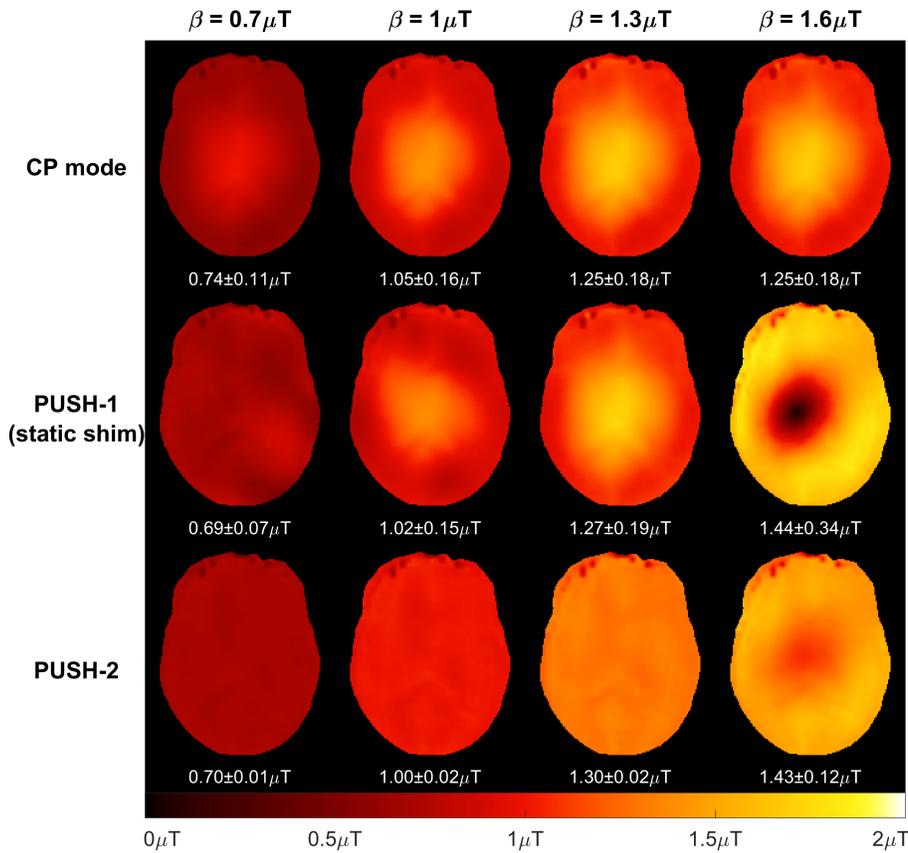

Supporting Information Figure S4: Corresponding 2D $B_1^{rms}$ maps for the MTR maps in Figure 6. Different rows correspond to different pulses (top row: CP mode; middle row: PUSH-1; bottom row: PUSH-2) and columns correspond to different $\beta$ (increasing from left to right). Below each is the mean $\pm$ standard deviation of $B_1^{rms}$ over the white matter mask also used in Figure 6.

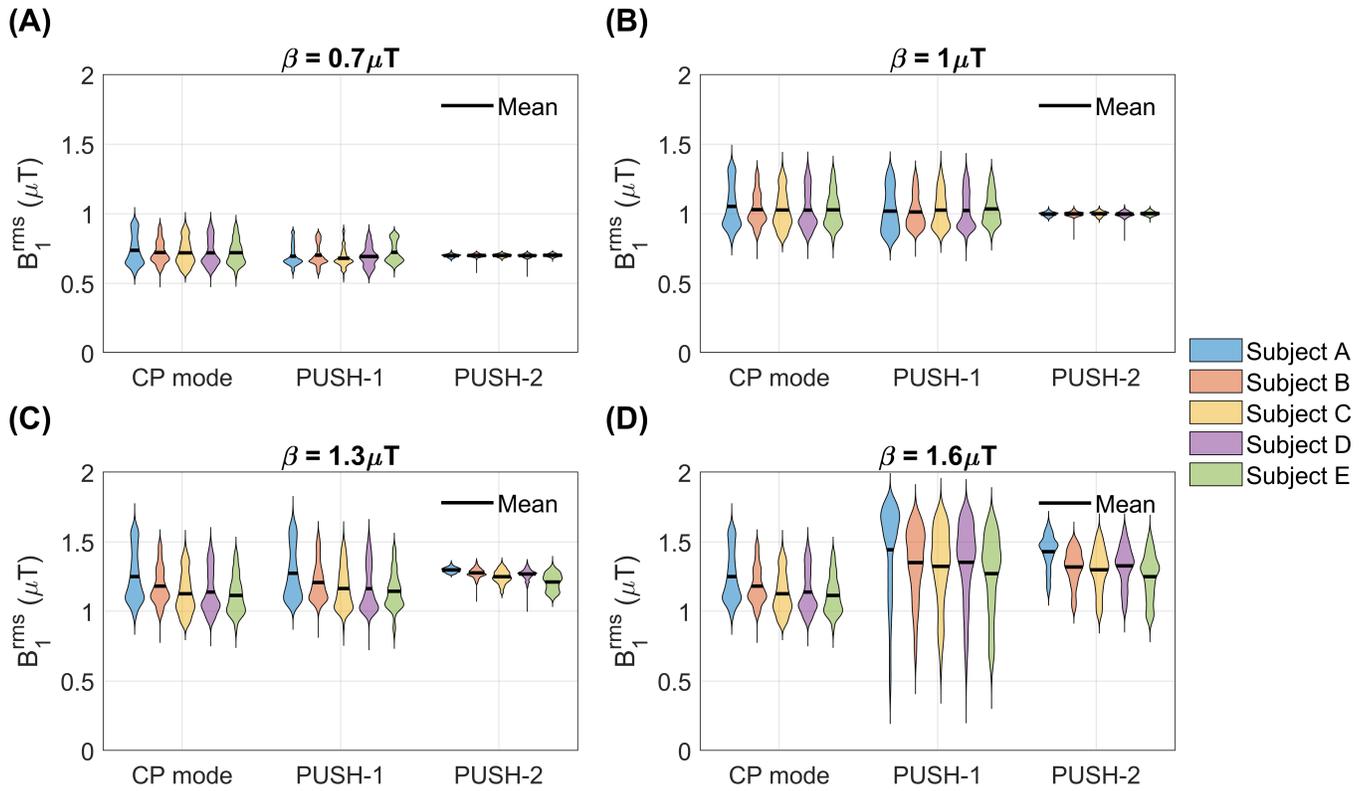

Supporting Information Figure S5: Corresponding $B_1^{rms}$ violin plot distributions for the MTR data in Figure 7. $B_1^{rms}$ distributions for saturation pre-pulses designed using $\beta$ of (A) $0.7\mu T$, (B) $1\mu T$, (C) $1.3\mu T$ and (D) $1.6\mu T$. The black line represents the mean $B_1^{rms}$. Subjects are sorted in increasing order of reference voltage, which is inversely proportional to the maximum $\beta$ achievable with CP mode.

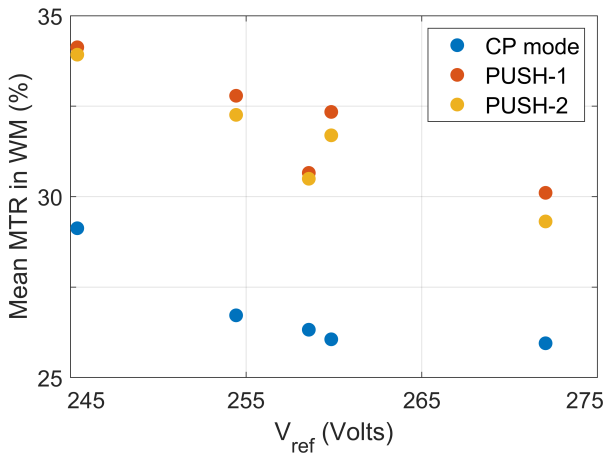

Supporting Information Figure S6: Mean MTR in WM ($\beta = 1.6\mu T$) versus the reference voltage $V_{ref}$ associated to each subject. At the highest $\beta$ all pulses are at the SAR limits and the mean MTR is indicative of the maximum MTR achieved.

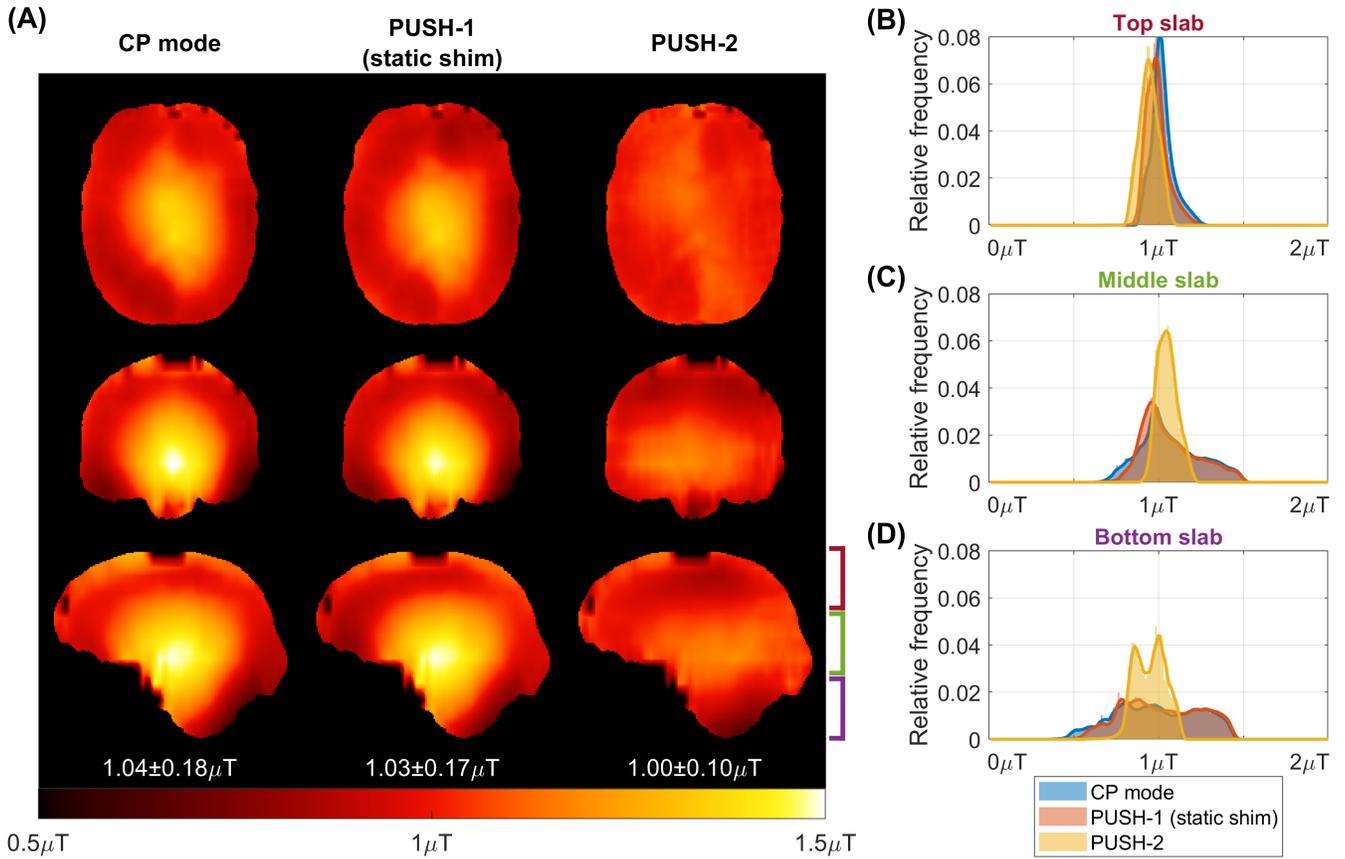

Supporting Information Figure S7: Corresponding $B_1^{\text{rms}}$ for the MTR data in Figure 8. (A) Transverse, coronal and sagittal planes of the 3D $B_1^{\text{rms}}$ maps. The left column contains the $B_1^{\text{rms}}$ maps using CP mode, middle column using PUSH-1 and right column using PUSH-2. Below each sagittal plane is the mean ± standard deviation of $B_1^{\text{rms}}$ over the white matter mask also used in Figure 8. (B-D) Histograms of the $B_1^{\text{rms}}$ distribution in white matter over three slabs: (B) top slab, (C) middle slab, and (D) bottom slab, as illustrated in (A) near the bottom right sagittal plane. Moving average plotted jointly with histograms to delineate distribution trend.

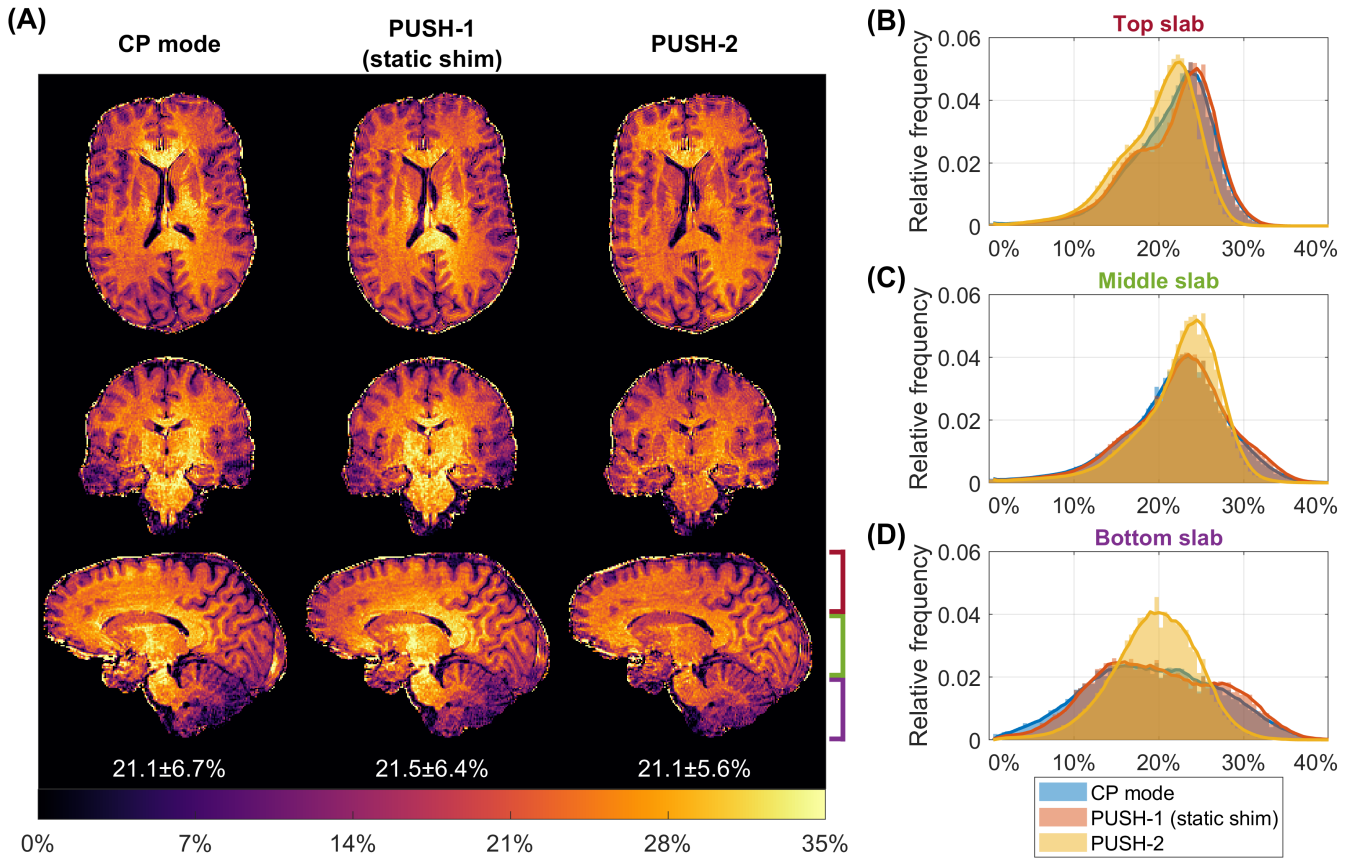

Supporting Information Figure S8: (A) Transverse, coronal and sagittal planes of the 3D MTR maps for subject E ($\beta = 1\mu T$). The left column contains the MTR maps acquired using CP mode, middle column using PUSH-1 and right column using PUSH-2. Below each sagittal plane is the mean $\pm$ standard deviation of MTR over the white matter mask. (B-D) Histograms of the MTR distribution in white matter over three slabs: (B) top slab, (C) middle slab, and (D) bottom slab, as illustrated in (A) near the bottom right sagittal plane. Moving average plotted jointly with histograms to delineate distribution trend.

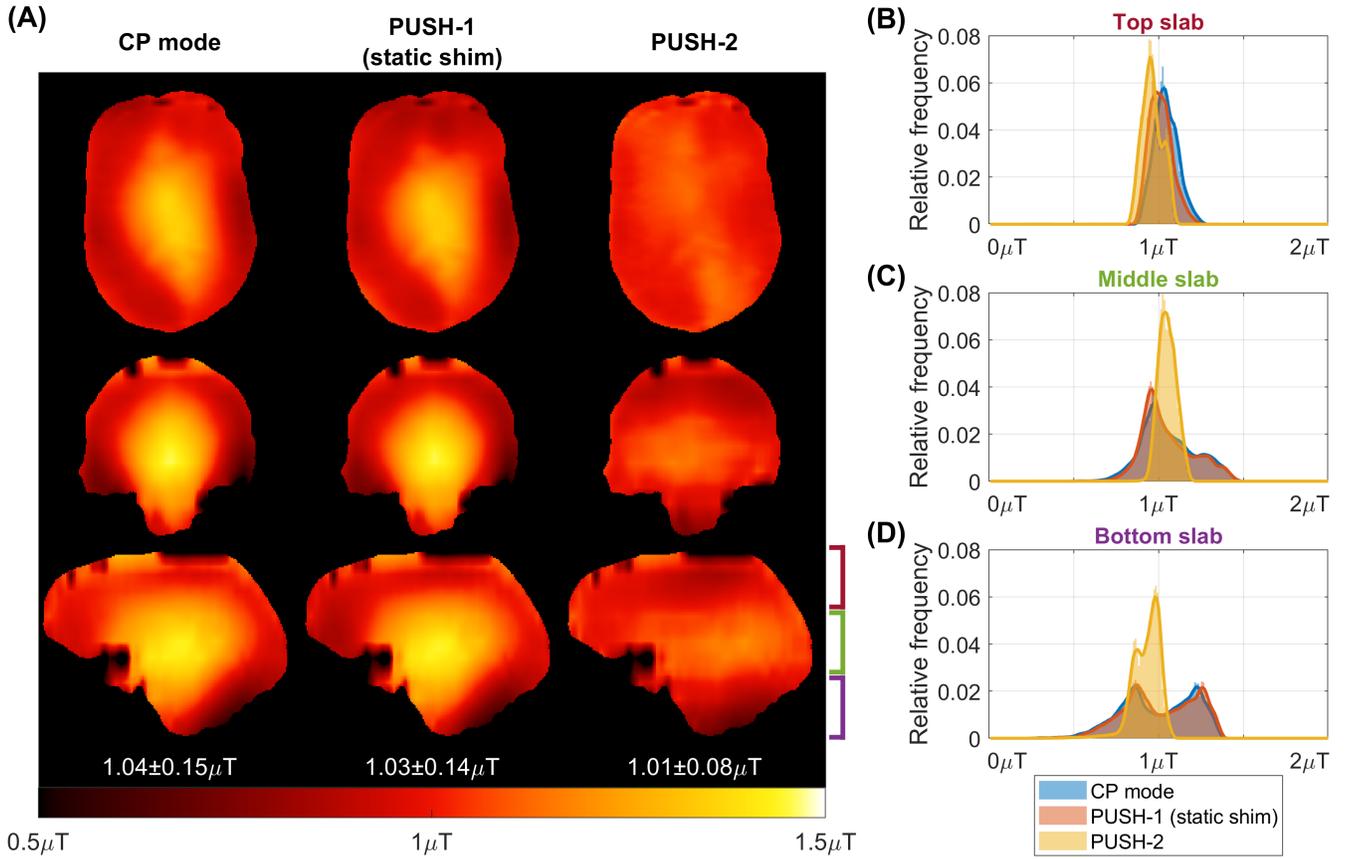

Supporting Information Figure S9: Corresponding $B_1^{rms}$ for the MTR data in Supporting Information Figure S8. (A) Transverse, coronal and sagittal planes of the 3D $B_1^{rms}$ maps. The left column contains the $B_1^{rms}$ maps using CP mode, middle column using PUSH-1 and right column using PUSH-2. Below each sagittal plane is the mean $\pm$ standard deviation of $B_1^{rms}$ over the white matter mask also used in Supporting Information Figure S8. (B-D) Histograms of the $B_1^{rms}$ distribution in white matter over three slabs: (B) top slab, (C) middle slab, and (D) bottom slab, as illustrated in (A) near the bottom right sagittal plane. Moving average plotted jointly with histograms to delineate distribution trend.